\documentclass{svjour2}                    
\smartqed  
\usepackage{graphicx}
\newcommand{\be}{\begin{eqnarray}}
\newcommand{\ee}{\end{eqnarray}}
\newcommand{\bn}{\begin{eqnarray*}}
\newcommand{\en}{\end{eqnarray*}}

\newcommand{\nn}{\nonumber \\}
\newcommand{\nl}{\\}

\renewcommand{\d}{\mbox{\rm d}}

\newcommand{\al}{\ensuremath{\alpha}}
\newcommand{\bt}{\ensuremath{\beta}}
\newcommand{\sg}{\ensuremath{\sigma}}
\newcommand{\gm}{\ensuremath{\gamma}}
\newcommand{\dl}{\ensuremath{\delta}}
\newcommand{\lm}{\ensuremath{\lambda}}

\newcommand{\Dl}{\ensuremath{\Delta}}

\newcommand{\Gm}{\ensuremath{\Gamma}}

\newcommand{\lt}{\ensuremath{\left}}
\newcommand{\rt}{\ensuremath{\right}}

\renewcommand{\d}{\mbox{\rm d}}

\begin{document}

\title{An Analytic Perturbation Approach for Classical Spinning Particle Dynamics
}


\author{Dinesh Singh
}


\institute{D. Singh \at
              Department of Physics \\ University of Regina \\ Regina, Saskatchewan \\ S4S 0A2, Canada \\ \\
              Tel.: +1-306-585-4681\\
              Fax: +1-306-585-5659\\ \\
              \email{singhd@uregina.ca}           
}

\date{Received: date / Accepted: date}

\maketitle

\begin{abstract}
A perturbation method to analytically describe the dynamics of a classical spinning particle,
based on the Mathisson-Papapetrou-Dixon (MPD) equations of motion, is presented.
By a power series expansion with respect to the particle's spin magnitude, it is shown how to obtain in general form
an analytic representation of the particle's kinematic and dynamical degrees of freedom that is formally applicable to
infinite order in the expansion.
Within this formalism, it is possible to identify a classical analogue of radiative corrections to the particle's mass and
spin due to spin-gravity interaction.
The robustness of this approach is demonstrated by showing how to explicitly compute the first-order momentum
and spin tensor components for arbitrary particle motion in a general space-time background.
Potentially interesting applications based on this perturbation approach are outlined.

\keywords{classical spinning particles \and spin-gravity interaction \and perturbation approach}
\end{abstract}

\section{Introduction}
\label{sec:1}

For many years, there has been on-going research into better understanding the motion of extended bodies in strong
gravitational fields, as described within the framework of general relativity.
Since it is reasonable to expect that such bodies will possess spin angular momentum during their formation,
it is critical to properly account for all interactions involving the coupling of their spin to the curved space-time background.
This understanding has relevance, for example, in identifying the motion of rapidly rotating neutron stars in orbit
around supermassive black holes, a topic of particular interest for the space-based LISA gravitational wave observatory \cite{LISA}
in observing low-frequency gravitational radiation emitted from these type of sources.

The first recorded attempt to solve this problem was formulated by Mathisson \cite{Mathisson}, who obtained an interaction
term in the form of a direct spin coupling to the Riemann curvature tensor.
Papapetrou \cite{Papapetrou} extended this initial idea by modelling the spinning particle as a matter field confined
within a space-time world tube.
Even more sophisticated approaches have been put forward by other people \cite{Tulczyjew},
primarily to deal with the higher-order multipole moment contributions to the spinning extended object's motion in curved space-time.
In certain respects, it can be argued that Dixon \cite{Dixon1,Dixon2} produced the most complete model to date,
which accounts for all multipole moment interactions to infinite order.
For most practical purposes, however, it is sufficient to truncate the equations of motion to leading order
in the spin, known as the ``pole-dipole approximation'' introduced by Mathisson and Papapetrou,
on the condition that the the spinning object's dimensions are small compared to the local radius of curvature
for the space-time background.
Within this approximation scheme, these equations are commonly referred to as the Mathisson-Papapetrou-Dixon (MPD) equations.

Besides a formal analysis of the MPD equations \cite{Ehlers,Bailey,Noonan}, there are particular studies on
spinning particle dynamics in a Kerr background \cite{Mashhoon1,Wald,Tod,Semerak}, scattering interactions with gravitational waves
\cite{Mohseni,Kessari}, and other applications.
In particular, the MPD equations lend themselves well to numerical analysis with applications involving gravitational wave
generation in a Kerr background \cite{Mino,Tanaka}, evidence of deterministic chaos within particle orbital
dynamics \cite{Suzuki1,Suzuki2,Hartl1,Hartl2}, and particle motion in a Vaidya background with radially infalling radiation \cite{Singh1}.
Recently, a first-order perturbative analysis of the MPD equations was developed by Chicone, Mashhoon, and Punsly (CMP) \cite{Chicone},
who applied their formalism to the study of rotating plasma clumps in astrophysical jets.
As well, Mashhoon and Singh \cite{Mashhoon2} applied the CMP approximation to the orbital dynamics of spinning particles in a Kerr
background, to compute leading order perturbations about a circular orbit and also explore the gravitomagnetic clock effect for spinning particles.

This paper presents a perturbation analysis of the MPD equations following the initial attempt made by CMP, but now extended to arbitrary
order in the perturbation expansion due to spin.
For what is presented below, geometric units of $G = c =1$ are assumed and the metric has $+2$ signature.
It begins with a brief outline of the MPD equations, followed by the perturbation approach adopted here.
Some formal calculations of relevant kinematic and dynamical quantities based on this approach are then presented,
leading to explicit calculation of the spinning particle's first-order momentum and spin tensor components for a particle
with unspecified motion in a general space-time background.
After discussion of some potential applications that may be relevant to this formalism, a brief conclusion follows.

\section{Mathisson-Papapetrou-Dixon (MPD) Equations}
\label{sec:2}

\paragraph{Equations of Motion:} The starting point for describing spinning particle motion in the ``pole-dipole approximation''
consists of the MPD equations of motion
\be
{DP^\mu \over \d \tau} & = & - {1 \over 2} \, R^\mu{}_{\nu \al \bt} \, u^\nu \, S^{\al \bt} \, ,
\label{MPD-momentum}
\nl
\nn
{DS^{\al \bt} \over \d \tau} & = & P^\al \, u^\bt - P^\bt \, u^\al \, ,
\label{MPD-spin}
\ee
where $P^\mu(\tau)$ is the spinning particle's linear momentum, $S^{\al \bt}(\tau)$ is the particle's antisymmetric spin angular momentum tensor,
$u^\mu(\tau) = \d x^\mu(\tau)/\d \tau$ is the particle's centre-of-mass four-velocity, and $R_{\mu \nu \al \bt}$ is the Riemann curvature tensor.
While $\tau$ will later become identified with the proper time of the particle along its centre-of mass worldline, within the context of
the MPD equations it is strictly just a parametrization variable whose properties need to be specified separately.
When considering more sophisticated models of spinning objects beyond the pole-dipole approximation, (\ref{MPD-momentum}) and (\ref{MPD-spin})
will have extra terms of the form $\cal F^\mu$ and $\cal T^{\al \bt}$ \cite{Ehlers,Mashhoon2} to denote additional forces and torques, respectively, based on
higher-order multipole moments beyond the mass monopole and spin dipole moment.
These extra terms require specification of the object's energy-momentum tensor $T^{\mu \nu}$ \cite{Dixon1,Dixon2,Mashhoon1,Mashhoon2}, subject to
covariant energy-momentum conservation $T^{\mu \nu}{}_{; \nu} = 0$, which requires detailed knowledge of the particle's internal structure.
However, for most practical purposes, the pole-dipole approximation is satisfactory.

\paragraph{Supplementary Equations:} By themselves, (\ref{MPD-momentum}) and (\ref{MPD-spin}) are insufficient to completely specify
the motion of a spinning particle in curved space-time.
To begin, the mass and spin parameters $m$ and $s$ are naturally identified in the form
\be
m^2 & = & -P_\mu \, P^\mu \, ,
\label{mass}
\nl
\nn
s^2 & = & {1 \over 2} \, S_{\mu \nu} \, S^{\mu \nu} \, .
\label{spin}
\ee
Furthermore, a supplementary spin condition needs to be specified in order to determine the particle's centre-of-mass trajectory.
This is best accomplished, following Dixon's approach \cite{Dixon1,Dixon2}, by imposing an orthogonality condition between the particle's linear
and spin angular momenta, in the form
\be
S^{\alpha \beta} \, P_\beta & = & 0 \, .
\label{spin-condition}
\ee
While both (\ref{mass}) and (\ref{spin}) are formally functions of $\tau$, it can easily be shown \cite{Chicone} using the MPD equations
and (\ref{spin-condition}) that $m$ and $s$ are {\em constants of the motion}.

With (\ref{spin-condition}) now specified, it is known \cite{Tod} that the four-velocity $u^\mu$ can be explicitly described in terms of
$P^\mu$ and $S^{\al \bt}$, such that
\be
u^\mu & = & -{P \cdot u \over m^2} \lt[P^\mu
+ {1 \over 2} \, {S^{\mu \nu} \, R_{\nu \gm \al \bt} \, P^\gm \, S^{\al \bt} \over
m^2 + {1 \over 4} \, R_{\al \bt \rho \sg} \, S^{\al \bt} \, S^{\rho \sg}} \rt],
\label{MPD-velocity}
\ee
where $P \cdot u$ is currently an undetermined quantity.
A value for this scalar product needs to be chosen in order to determine the rate of the particle's internal clock with respect to $\tau$.
It is important to emphasize from (\ref{MPD-velocity}) that the linear momentum and four-velocity are {\em not} co-linear due to a
non-trivial spin-curvature interaction.
This has the effect of pulling the particle off a geodesic worldline, leading to interesting consequences for understanding the
interplay between the particle's centre-of-mass motion and the dynamical response generated by the spin interaction with space-time curvature.

\section{Perturbation Approach to MPD Equations}
\label{sec:3}

\subsection{CMP Approximation}
\label{sec:3.1}

As noted earlier, a first attempt in describing the MPD equations as a perturbation expansion was performed \cite{Chicone,Mashhoon2},
in which the underlying assumption is that $P^\mu - m \, u^\mu = E^\mu$ is small, where $E^\mu$ denotes the spin-curvature force.
Furthermore, the spin magnitude is such that the M{\o}ller radius $\rho$ \cite{Chicone,Mashhoon2,Moller} is $\rho = s/m \ll r$, where $r$ is the distance
from the particle to the source generating space-time curvature.
This leads to the CMP approximation, expressed in the form
\be
{D P^\mu \over \d \tau} & \approx & -{1 \over 2} \, R^\mu{}_{\nu \al \bt} \, u^\nu \, S^{\al \bt}\, ,
\label{CMP-momentum}
\nl
\nn
{DS^{\mu \nu} \over \d \tau} & \approx & 0 \, ,
\label{CMP-spin}
\ee
where the associated spin condition is
\be
S_{\mu \nu} \, u^\nu & \approx & 0 \, .
\label{CMP-spin-condition}
\ee

It is important to recognize from (\ref{CMP-spin}) that, to first-order in $s$, the spin tensor is parallel transported in space-time.
When comparing the CMP approximation with a numerical integration of the MPD equations for circular motion around a Kerr black hole \cite{Mashhoon2},
it is shown that the kinematic properties of the spinning particle agree very well in general.
However, there is some loss of agreement when considering the radial component of the particle's motion, since the CMP approximation
does not reveal any modulation of its radial position compared to the corresponding plots generated by the MPD equations.
This discrepancy appears when $s/(mr) \sim 10^{-2}-10^{-1}$ for $r = 10 \, M$, where $M$ is the mass of the Kerr black hole.
It is most likely due to the lack of a more complicated spin interaction beyond what can be generated by (\ref{CMP-spin}).
With this in mind, it seems appropriate to consider whether a more detailed approximation can account for the extra structure
missing in the radial plots based on the CMP approximation alone, which is part of the motivation for this investigation.

\subsection{Formalism}
\label{sec:3.2}

\paragraph{Equations of Motion:} The underlying assumption within this formalism is to define the particle's linear momentum and spin angular momentum in the form
\be
P^\mu(\varepsilon) & \equiv & \sum_{j = 0}^\infty \varepsilon^j \, P_{(j)}^\mu \, , 
\label{P-approx-def}
\nl
\nn
S^{\mu \nu} (\varepsilon) & \equiv & \varepsilon \sum_{j = 0}^\infty \varepsilon^j \, S_{(j)}^{\mu \nu} \, ,
\label{S-approx-def}
\ee
where $\varepsilon$ is an expansion parameter corresponding to powers of $s$ present in each order of the expansion,
and $P_{(j)}^\mu$ and $S_{(j)}^{\mu \nu}$ are the jth-order contributions of the linear momentum and spin angular momentum, respectively.
That is, the zeroth-order quantities in $\varepsilon$ correspond to the dynamics of a spinless particle,
while the first-order quantities in $\varepsilon$ lead to the CMP approximation described by (\ref{CMP-momentum}) and (\ref{CMP-spin}).
This is confirmed when substituting (\ref{P-approx-def}) and (\ref{S-approx-def}) into
\be
{DP^\mu (\varepsilon) \over \d \tau} & = & -{1 \over 2} \, R^\mu{}_{\nu \al \bt} \, u^\nu (\varepsilon) \, S^{\al \bt} (\varepsilon) \, ,
\label{MPD-momentum-e}
\nl
\nn
{DS^{\al \bt}(\varepsilon) \over \d \tau} & = &  2 \, \varepsilon \, P^{[\al}(\varepsilon) \, u^{\bt]}(\varepsilon) \, ,
\label{MPD-spin-e}
\ee
where for (\ref{MPD-spin-e}) the square brackets denote antisymmetrization of the indices defined by
$A^{[\al} \, B^{\bt]} \equiv {1 \over 2} \lt(A^\al \, B^\bt - A^\bt \, B^\al\rt)$,
and an extra factor of $\varepsilon$ has to be added to maintain consistency with the definition (\ref{S-approx-def}).
Furthermore,
\be
u^\mu(\varepsilon) & \equiv & \sum_{j = 0}^\infty \varepsilon^j \, u_{(j)}^\mu \, ,
\label{MPD-velocity-e}
\ee
where the explicit expressions for $u_{(j)}^\mu$ are determined separately.
When (\ref{P-approx-def})--(\ref{MPD-velocity-e}) are combined together and grouped in terms of like orders of $\varepsilon$, it follows that
the jth-order expressions of the MPD equations are
\be
{DP_{(j)}^\mu \over \d \tau} & = & - \frac{1}{2} \, R^\mu{}_{\nu \alpha \beta} \sum_{k = 0}^{j - 1} u_{(j-1-k)}^\nu \, S_{(k)}^{\alpha \beta} \, ,
\label{DP-j}
\nl
\nn
{DS_{(j)}^{\alpha \beta} \over \d \tau} & = &  2 \sum_{k = 0}^j P_{(j-k)}^{[\alpha} \, u_{(k)}^{\beta]} \, ,
\label{DS-j}
\ee
where, for $P_{(0)}^\mu = m_0 \, u_{(0)}^\mu$ with
\be
m_0^2 & \equiv & -P^{(0)}_\mu \, P_{(0)}^\mu \, ,
\label{m0-sq}
\ee
the zeroth- and first-order terms of (\ref{DP-j}) and (\ref{DS-j}) in $\varepsilon$ are
\be
{D P_{(0)}^\mu \over \d \tau} & = & 0 \, ,
\label{DP-0}
\ee
and
\be
{D P_{(1)}^\mu \over \d \tau} & = & -{1 \over 2} \, R^\mu{}_{\nu \al \bt} \, u_{(0)}^\nu \, S_{(0)}^{\al \bt} \, ,
\label{DP-1}
\nl
\nn
{D S_{(0)}^{\mu \nu} \over \d \tau} & = & 0 \, .
\label{DS-0}
\ee
This is consistent with the main results (\ref{CMP-momentum}) and (\ref{CMP-spin}) of the CMP approximation.

\paragraph{Perturbation of the Mass and Spin Parameters:} Besides the MPD equations, the supplementary equations
(\ref{mass})--(\ref{MPD-velocity}) serve a very important function in the development of this perturbation approach.
This is especially true concerning the mass and spin magnitude parameters $m$ and $s$, since they are dependent on
$P^\mu$ and $S^{\mu \nu}$, which are then expressible in terms of (\ref{P-approx-def}) and (\ref{S-approx-def}).
In fact, by invoking the language of quantum field theory, it becomes possible to identify the classical
analogue of a {\em bare mass} $m_0$ defined according to (\ref{m0-sq}), along with a {\em bare spin} $s_0$, where
\be
s_0^2 & \equiv & {1 \over 2} \, S_{\mu \nu}^{(0)} \, S_{(0)}^{\mu \nu} \, ,
\label{s0-sq}
\ee
such that the total mass and spin magnitudes exist as the sum of ``radiative corrections'' to $m_0$ and $s_0$,
due to the MPD equations in perturbative form.
That is,
\be
m^2 (\varepsilon) & = & m_0^2 \lt(1 + \sum_{j=1}^\infty \varepsilon^j \, \bar{m}_j^2 \rt),
\label{m-sq}
\nl
\nn
s^2 (\varepsilon) & = & \varepsilon^2 \, s_0^2 \lt(1 + \sum_{j=1}^\infty \varepsilon^j \, \bar{s}_j^2 \rt),
\label{s-sq}
\ee
where $\bar{m}_j$ and $\bar{s}_j$ are dimensionless jth-order corrections to $m_0$ and $s_0$, respectively,
in the form
\be
\bar{m}_j^2 & = & - {1 \over m_0^2} \, \sum_{k=0}^j P_\mu^{(j-k)} \, P_{(k)}^\mu \, ,
\label{m-bar-sq}
\nl
\nn
\bar{s}_j^2 & = & {1 \over s_0^2} \, \sum_{k=0}^j S_{\mu \nu}^{(j-k)} \, S_{(k)}^{\mu \nu} \, .
\label{s-bar-sq}
\ee
Explicit expressions of (\ref{m-bar-sq}) and (\ref{s-bar-sq}) required for subsequent calculations are
\be
\bar{m}_1^2 & = & -{2 \over m_0^2} \, P_\mu^{(1)} \, P_{(0)}^\mu \, ,
\label{m-bar1}
\nl
\nn
\bar{m}_2^2 & = & -{1 \over m_0^2} \lt[2 \, P_\mu^{(2)} \, P_{(0)}^\mu + P_\mu^{(1)} \, P_{(1)}^\mu \rt],
\label{m-bar2}
\nl
\nn
\bar{m}_3^2 & = & -{2 \over m_0^2} \lt[P_\mu^{(3)} \, P_{(0)}^\mu + P_\mu^{(2)} \, P_{(1)}^\mu \rt],
\label{m-bar3}
\ee
and
\be
\bar{s}_1^2 & = & {2 \over s_0^2} \, S_{\mu \nu}^{(1)} \, S_{(0)}^{\mu \nu} \, ,
\label{s-bar1}
\nl
\nn
\bar{s}_2^2 & = & {1 \over s_0^2} \lt[2 \, S_{\mu \nu}^{(2)} \, S_{(0)}^{\mu \nu} + S_{\mu \nu}^{(1)} \, S_{(1)}^{\mu \nu}\rt] .
\label{s-bar2}
\ee
Since $m$ and $s$ are already known to be constant for an exact treatment of the MPD equations,
it follows that $\bar{m}_j$ and $\bar{s}_j$ must also be constant for each order of the perturbation expansion in $\varepsilon$.
While it is not obvious that this property emerges from (\ref{m-bar-sq}) and (\ref{s-bar-sq}),
this can be verified to at least the orders of expansion considered in this paper, upon evaluating the perturbative form of $u^\mu$.

\subsection{Kinematic and Dynamical Quantities}
\label{sec:3.3}

\paragraph{Four-Velocity:} As noted earlier, supplementary equations are needed to completely specify the motion of a spinning particle
according to the MPD equations.
This includes the four-velocity described by (\ref{MPD-velocity}) once the undetermined scalar product $P \cdot u$ is specified.
It turns out that by setting
\be
P \cdot u & \equiv & -m(\varepsilon),
\label{P.u}
\ee
it follows that
\be
u^\mu(\varepsilon) 
& = & {1 \over m(\varepsilon)} \left[P^\mu(\varepsilon)
+ {1 \over 2} \, {S^{\mu \nu}(\varepsilon) \, R_{\nu \gamma \alpha \beta} \, P^\gamma(\varepsilon) \, S^{\alpha \beta}(\varepsilon) \over m^2(\varepsilon) \, \Dl(\varepsilon)}
\right],
\label{MPD-velocity-fixed}
\ee
where
\be
\Dl(\varepsilon) & \equiv & 1 + {1 \over 4 \, m^2(\varepsilon)} \, R_{\mu \nu \al \bt} \, S^{\mu \nu} (\varepsilon) \, S^{\al \bt} (\varepsilon) \, .
\label{Delta}
\ee
With this particular choice of parametrization constraint, it is straightforward to see that (\ref{MPD-velocity-fixed}),
along with the spin condition constraint equation (\ref{spin-condition}), leads to
\be
u_\mu(\varepsilon) \, u^\mu(\varepsilon) & = & -1 + {1 \over 4 \, m^6(\varepsilon) \, \Dl^2 (\varepsilon)}
\, \tilde{R}_\mu(\varepsilon) \, \tilde{R}^\mu (\varepsilon) \ = \ -1 + O(\varepsilon^4) \, ,
\label{u.u}
\ee
where
\be
\tilde{R}^\mu(\varepsilon) & \equiv & S^{\mu \nu}(\varepsilon) R_{\nu \gm \al \bt} \, P^\gm (\varepsilon) \, S^{\al \bt} (\varepsilon) \, .
\label{R-tilde}
\ee
At least to third-order in $\varepsilon$, this expression for (\ref{u.u}) assuming (\ref{P.u}) justifies the labelling of
$\tau$ as {\em proper time} for parameterizing the spinning particle's motion along its centre-of-mass worldline.

Though somewhat tedious to calculate, it is a straightforward exercise to evaluate the four-velocity $u^\mu$ perturbatively,
based on (\ref{MPD-velocity-fixed}) and using (\ref{P-approx-def}), (\ref{S-approx-def}), and (\ref{m-sq}).
Therefore, it follows that the explicit expression for the spinning particle's four-velocity in general form is
\be
u^\mu(\varepsilon) & = & \sum_{j = 0}^\infty \varepsilon^j \, u_{(j)}^\mu
\ = \ {P_{(0)}^\mu \over m_0} + \varepsilon \lt[{1 \over m_0} \lt(P_{(1)}^\mu - {1 \over 2} \, \bar{m}_1^2 \, P_{(0)}^\mu\rt) \rt]
\nn
\nn
\nn
&   & {} + \varepsilon^2 \lt\{ {1 \over m_0} \lt[P_{(2)}^\mu - {1 \over 2} \, \bar{m}_1^2 \, P_{(1)}^\mu
         - {1 \over 2} \lt(\bar{m}_2^2 - {3 \over 4} \, \bar{m}_1^4\rt) P_{(0)}^\mu \rt] \rt.
\nn
\nn
\nn
&   & {} + \lt. {1 \over 2 m_0^3} \, S_{(0)}^{\mu \nu} \, R_{\nu \gm \al \bt}  \, P_{(0)}^\gm \, S_{(0)}^{\al \bt} \rt\}
\nn
\nn
\nn
&   & {} + \varepsilon^3 \lt\{ {1 \over m_0} \lt[P_{(3)}^\mu - {1 \over 2} \, \bar{m}_1^2 \, P_{(2)}^\mu
         - {1 \over 2} \lt(\bar{m}_2^2 - {3 \over 4} \, \bar{m}_1^4\rt) P_{(1)}^\mu \rt. \rt.
\nn
\nn
\nn
&   & {} - \lt. {1 \over 2} \lt(\bar{m}_3^2 - {3 \over 2} \, \bar{m}_1^2 \, \bar{m}_2^2 + {5 \over 8} \, \bar{m}_1^6 \rt) P_{(0)}^\mu \rt]
\nn
\nn
&   & {} + \lt.
{1 \over 2 m_0^3} \, R_{\nu \gm \al \bt} \lt[\sum_{n=0}^1 S_{(1-n)}^{\mu \nu} \sum_{k=0}^n P_{(n-k)}^\gm \, S_{(k)}^{\al \bt}
- {3 \over 2} \, \bar{m}_1^2 \, S_{(0)}^{\mu \nu} \, P_{(0)}^\gm \, S_{(0)}^{\al \bt} \rt] \rt\}
\nn
\nn
\nn
&   & {} + O(\varepsilon^4) \, .
\label{MPD-velocity-explicit}
\ee
Given (\ref{m-bar1})--(\ref{m-bar3}), along with the spin condition constraint equation (\ref{spin-condition}) in the form
\be
P_\mu^{(0)} \, S_{(j)}^{\mu \nu} & = & - \sum_{k=1}^j P_\mu^{(k)} \, S_{(j-k)}^{\mu \nu} \, , \qquad j \geq 1
\label{s.p=0}
\ee
for the (j+1)th-order contribution in $\varepsilon$, with
\be
P_\mu^{(0)} \, S_{(0)}^{\mu \nu} & = & 0 \, ,
\label{s.p=0-e1}
\nl
\nn
P_\mu^{(0)} \, S_{(1)}^{\mu \nu} + P_\mu^{(1)} \, S_{(0)}^{\mu \nu} & = & 0 \, ,
\label{s.p=0-e2}
\ee
for the first- and second-order perturbations in $\varepsilon$, respectively, it is straightforward to verify that
(\ref{MPD-velocity-explicit}) satisfies (\ref{u.u}) to third order in $\varepsilon$.
It follows that the perturbation of the spinning particle's worldline away from a geodesic is attainable
by integration of (\ref{MPD-velocity-explicit}) with respect to $\tau$.

\paragraph{Constancy of the Mass and Spin Parameters:}
With (\ref{MPD-velocity-explicit}) evaluated, it is possible to verify that
\be
{D \bar{s}_j^2 \over \d \tau} & = & {D \bar{m}_j^2 \over \d \tau} \ = \ 0 \,
\label{Ds/dt-Dm/dt=0}
\ee
for at least (\ref{m-bar1})--(\ref{s-bar2}), given (\ref{DP-j}) and (\ref{DS-j}).
In particular, from evaluating $D \bar{m}_j^2/\d \tau$ directly according to (\ref{DP-j}), it is found that
\be
0 & = & R_{\mu \nu \al \bt} \lt(\sum_{l=1}^{j+2} P_{(j+2-l)}^\mu \sum_{k=0}^{l-1} u_{(l-1-k)}^\nu \rt) S_{(k)}^{\al \bt} \, ,
\label{Dmj-constraint}
\ee
which is identically satisfied for the required expressions of $u_{(j)}^\mu$ given by (\ref{MPD-velocity-explicit}).
Furthermore, it can be shown explicitly that the first- and second-order perturbations of the spin tensor satisfy
\be
{D S_{(1)}^{\mu \nu} \over \d \tau} & = & 0 \, ,
\label{DS1/dt=0}
\nl
\nn
{D S_{(2)}^{\mu \nu} \over \d \tau} & = & {1 \over m_0^3} \, P_{(0)}^{[\mu} \, S_{(0)}^{\nu]\sg} \, R_{\sg \gm \al \bt} \, P_{(0)}^\gm \, S_{(0)}^{\al \bt} \, .
\label{DS2/dt=0}
\ee

\paragraph{M{\o}ller Radius:} An important quantity to evaluate via this perturbation approach is
the M{\o}ller radius $\rho = s/m$, which has significance in determining, for example, the strength of
the spin-curvature force when applied to circular orbits of spinning particles in black hole space-times \cite{Singh1,Mashhoon2}.
Getting a better detailed sense of how the M{\o}ller radius appears due to this perturbation approach
may become useful for understanding the precise conditions for a transition into chaotic dynamics,
as suggested in previous work \cite{Suzuki1,Suzuki2,Hartl1,Hartl2}.
Given (\ref{m-bar1})--(\ref{s-bar2}), it is a straightforward calculation to show that
\be
{s(\varepsilon) \over m(\varepsilon)}
& = & {s_0 \over m_0} \lt\{\varepsilon + \varepsilon^2 \lt[{1 \over 2} \lt(\bar{s}_1^2 - \bar{m}_1^2\rt) \rt] \rt.
\nn
\nn
\nn
&  & \lt. {} + \varepsilon^3 \lt[{1 \over 2} \lt(\bar{s}_2^2 - \bar{m}_2^2\rt) - {1 \over 4} \, \bar{s}_1^2 \, \bar{m}_1^2
- {1 \over 8} \lt(\bar{s}_1^4 - 3 \, \bar{m}_1^4\rt)\rt] + O(\varepsilon^4) \rt\},
\label{s/m}
\ee
where non-trivial deviations from $\rho_0 = s_0/m_0$ due to geodesic motion emerge at second order in $\varepsilon$.
Since the ``radiative corrections'' to $m_0$ and $s_0$ described by (\ref{m-bar-sq}) and (\ref{s-bar-sq}) have the effect
of increasing both the particle's mass and spin simultaneously, the M{\o}ller radius appears to remain in the vicinity of $\rho_0$.
At present, however, it is impossible to precisely determine the nature of the shift from $\rho_0$ to $\rho$
without first examining (\ref{s/m}) with respect to a particular space-time background.

\subsection{First-Order Linear Momentum and Spin Components}
\label{sec:3.4}

\paragraph{Local Fermi Co-ordinate System:}
To fully appreciate the value of this perturbation approach for describing the MPD equations,
it is useful to compute the first-order linear momentum and spin angular momentum components in general form.
Evaluation of $P_{(1)}^\mu$ is particularly straightforward when formulating the problem in terms
of a local Fermi co-ordinate system \cite{Mashhoon2} and an orthonormal tetrad frame $\lm^\mu{}_{\hat{\al}}$.
This leads to $P_{(1)}^\mu = \lm^\mu{}_{\hat{\al}} \, P_{(1)}^{\hat{\al}}$, where $\hat{\al}$ denote indices
for the Fermi co-ordinates $x^{\hat{\al}}$ defined on a locally flat tangent space in the neighbourhood of the
spinning particle, and $P_{(1)}^{\hat{\al}}$ is the corresponding expression for the linear momentum on the tangent space.
The tetrad frame, with $\lm^\mu{}_{\hat{0}} = u_{(0)}^\mu$, satisfies the orthonormality condition
\be
\eta_{\hat{\al} \hat{\bt}} & = & g_{\mu \nu} \, \lm^\mu{}_{\hat{\al}} \, \lm^\nu{}_{\hat{\bt}} \, ,
\label{ortho-tetrad}
\ee
and satisfies the parallel transport law
\be
{D \lm^\mu{}_{\hat{\al}} \over \d \tau} & = & 0 \,
\label{tetrad-parallel-transport}
\ee
with respect to the general space-time co-ordinates $x^\mu$.
The Riemann curvature tensor projected onto the locally flat tangent space then satisfies
\be
{}^F{}R_{\hat{\al} \hat{\bt} \hat{\gm} \hat{\dl}} & = & R_{\mu \nu \rho \sg} \,
\lm^\mu{}_{\hat{\al}} \, \lm^\nu{}_{\hat{\bt}} \, \lm^\rho{}_{\hat{\gm}} \, \lm^\sg{}_{\hat{\dl}} \, .
\label{F-Riemann}
\ee
In addition, the first-order spin condition (\ref{s.p=0-e1}) with (\ref{ortho-tetrad}) requires that
$S_{(0)}^{\mu \nu} = \lm^\mu{}_{\hat{\imath}} \, \lm^\nu{}_{\hat{\jmath}} \, S_{(0)}^{\hat{\imath} \hat{\jmath}}$
to preserve orthogonality with $P_{(0)}^\mu$ in terms of the tetrad frame \cite{Mashhoon2}.

\paragraph{First-Order Linear Momentum:}
Given the tetrad formalism presented here, evaluation of the first-order linear momentum components becomes
very easy to implement.
From (\ref{DP-1}) and (\ref{tetrad-parallel-transport}), it follows that
$D P_{(1)}^\mu / \d \tau = \lm^\mu{}_{\hat{\al}} \lt(\d P_{(1)}^{\hat{\al}} / \d \tau \rt)$, which results in
\be
{\d P_{(1)}^{\hat{\al}} \over \d \tau} & = & -{1 \over 2} \, {}^F{}R^{\hat{\al}}{}_{\hat{0} \hat{\imath} \hat{\jmath}} \, S_{(0)}^{\hat{\imath} \hat{\jmath}} \,
\label{dP/dt}
\ee
to be solved.
This leads to the general expression
\be
P_{(1)}^\mu & = & -{1 \over 2} \, \lm^\mu{}_{\hat{\al}}
\int \lt({}^F{}R^{\hat{\al}}{}_{\hat{0} \hat{\imath} \hat{\jmath}} \, S_{(0)}^{\hat{\imath} \hat{\jmath}}\rt) \d \tau \, .
\label{P1-general}
\ee

\paragraph{First-Order Spin Angular Momentum:}
Determining the first-order spin tensor components, in contrast to that of the linear momentum, is not so straightforward.
This is because a similar line of reasoning when applied to (\ref{DS1/dt=0}) leads to the expression
\be
{D S_{(1)}^{\mu \nu} \over \d \tau} & = & \lm^\mu{}_{\hat{\al}} \, \lm^\nu{}_{\hat{\bt}} \, {\d S_{(1)}^{\hat{\al} \hat{\bt}} \over \d \tau} \ = \ 0 \, ,
\label{DS1/dt=0-tetrad}
\ee
in which the $\d S_{(1)}^{\hat{\al} \hat{\bt}}/\d \tau$ defined on the local tangent space are still {\em undetermined}.
Specifically, it is {\em not} necessarily true that $\d S_{(1)}^{\hat{\al} \hat{\bt}}/\d \tau = 0$, since the six
equations in (\ref{DS1/dt=0-tetrad}) may not all be linearly independent of each other.
Furthermore, even if this requirement were to be satisfied, the resulting constants of integration $S_{(1)}^{\hat{\al} \hat{\bt}}$
cannot be further specified without some physical justification.
Therefore, a different approach is required.

Fortunately, there is a means available to solve for $S_{(1)}^{\mu \nu}$ in general form, based on making full use of
the spin condition constraint equation (\ref{spin-condition}).
Given (\ref{s.p=0-e2}), there are four equations of the form
\be
A_\mu \, S_{(1)}^{\mu \nu} - B^\nu & = & 0 \, ,
\label{S1-4-equations}
\ee
where
\be
A_\mu & \equiv & P_\mu^{(0)} \, ,
\label{A}
\nl
\nn
B^\nu & \equiv & -P_\mu^{(1)} \, S_{(0)}^{\mu \nu} \, .
\label{B}
\ee
As well, contraction of (\ref{s.p=0-e2}) with $P_\nu^{(1)}$ leads to two constraint equations of the form
\be
C_{\mu \nu} \, S_{(1)}^{\mu \nu} & = & 0 \, ,
\label{C-constraint}
\nl
\nn
E_{\mu \nu} \, S_{(1)}^{\mu \nu} & = & 0 \, ,
\label{E-constraint}
\ee
where
\be
C_{\mu \nu} & \equiv & P^{(0)}_{[\mu} \, P^{(1)}_{\nu]} \, ,
\label{C}
\nl
\nn
E_{\mu \nu} & \equiv & P^{(0)}_{[\mu} \, D P^{(1)}_{\nu]}/ \d \tau \, ,
\label{E}
\ee
given (\ref{DP-0}) and (\ref{DS1/dt=0}).

With (\ref{S1-4-equations}), (\ref{C-constraint}), and (\ref{E-constraint}), there are six linear equations in $S_{(1)}^{\mu \nu}$
and six unknowns for the spin tensor components, so it should be possible to solve for a unique solution of $S_{(1)}^{\mu \nu}$ algebraically.
However, there is a complication in that one of the equations in (\ref{S1-4-equations}) is linearly {\em dependent} on the
other three, leading to the first-order spin condition constraint equation (\ref{s.p=0-e1}).
Therefore, the linear system of equations is formally underdetermined in one variable, leading to the solution
\be
S_{(1)}^{13} & = & - \lt(H_{12} \, J_{23} - J_{12} \, H_{23} \over H_{12} \, J_{13} - J_{12} \, H_{13}\rt) S_{(1)}^{23}
+ {1 \over A_0} \lt(J_{12} \, C_{0j} - H_{12} \, E_{0j} \over H_{12} \, J_{13} - J_{12} \, H_{13}\rt) B^j \, ,
\label{S13}
\nl
\nn
S_{(1)}^{12} & = & - \lt(H_{13} \over H_{12}\rt)S_{(1)}^{13} - \lt(H_{23} \over H_{12}\rt)S_{(1)}^{23} - {C_{0j} \, B^j \over A_0 \, H_{12}} \, ,
\label{S12}
\nl
\nn
S_{(1)}^{03} & = & - \lt(A_1 \over A_0\rt)S_{(1)}^{13} - \lt(A_2 \over A_0\rt)S_{(1)}^{23} + {B^3 \over A_0} \, ,
\label{S03}
\nl
\nn
S_{(1)}^{02} & = & - \lt(A_1 \over A_0\rt)S_{(1)}^{12} + \lt(A_3 \over A_0\rt)S_{(1)}^{23} + {B^2 \over A_0} \, ,
\label{S02}
\nl
\nn
S_{(1)}^{01} & = & \lt(A_2 \over A_0\rt)S_{(1)}^{12} + \lt(A_3 \over A_0\rt)S_{(1)}^{13} + {B^1 \over A_0} \, ,
\label{S01}
\ee
expressed in terms of $S_{(1)}^{23}$, and
\be
H_{ij} & = & C_{ij} + 2 \, C_{0[i} \, A_{j]}/A_0 \, ,
\label{H}
\nl
\nn
J_{ij} & = & E_{ij} + 2 \, E_{0[i} \, A_{j]}/A_0 \, .
\label{J}
\ee
The expressions (\ref{S13})--(\ref{S01}) can then be put into the form
\be
S_{(1)}^{\mu \nu} & = & M^{\mu \nu} \, S_{(1)}^{23} + N^{\mu \nu} \, ,
\label{S1-mn}
\ee
where $M^{\mu \nu}(\tau)$ and $N^{\mu \nu}(\tau)$ are antisymmetric in their indices, and
\be
M^{23} & = & 1 \, , \qquad N^{23} \ = \ 0 \, .
\label{M-N-23}
\ee

There is, however, one remaining equation to consider, namely (\ref{DS1/dt=0}) which
describes the parallel transport of $S_{(1)}^{\mu \nu}$.
Since this is true for every component of the first-order spin tensor, it must certainly apply for the case of $S_{(1)}^{23}$.
Therefore, given (\ref{DS1/dt=0}) and (\ref{S1-mn}), it follows that the first-order linear differential equation
\be
{\d S_{(1)}^{23}(\tau) \over \d \tau} + P(\tau) \, S_{(1)}^{23}(\tau) & = & Q(\tau) \,
\label{dS1-23/dt}
\ee
must be satisfied, where
\be
P(\tau) & = & -2 \, u_{(0)}^\al \, \Gm_{\al \bt}^{[2} \, M^{3]\bt} \, ,
\label{P}
\nl
\nn
Q(\tau) & = & 2 \, u_{(0)}^\al \, \Gm_{\al \bt}^{[2} \, N^{3]\bt} \, ,
\label{Q}
\ee
in terms of the metric connection $\Gm^\gm_{\al \bt}$.
Since the solution to (\ref{dS1-23/dt}) is known exactly  \cite{Nagle} in terms of a suitably chosen integrating factor $\mu(\tau)$,
it follows that
\be
S_{(1)}^{23}(\tau) & = & {1 \over \mu(\tau)} \lt(\int_0^\tau \mu(\tau') \, Q(\tau') \, \d \tau'\rt) \, ,
\label{S1-23}
\ee
where
\be
\mu(\tau) & = & \exp \lt(\int_0^\tau P(\tau') \, \d \tau' \rt) ,
\label{mu}
\ee
and $S_{(1)}^{23}(0) = 0$.
Given the generally non-trivial dependence of $\tau$ on the integrands in (\ref{S1-23}) and (\ref{mu}),
it is most likely that $S_{(1)}^{23}$ will have to be solved numerically.
While presented in a compact form, it is rather surprising that the solution (\ref{S1-23}) possesses a considerable
level of complexity.
It will be interesting to analyze the structure of $S_{(1)}^{\mu \nu}$ when calculated for particle motion in a simple but non-trivial
space-time background, such as the Schwarzschild metric.

\section{Potential Applications}
\label{sec:4}

It seems clear that this perturbation approach to the MPD equations leads to a robust formalism that
lends itself well to various applications.
For example, following upon previous work \cite{Mashhoon2}, it is possible to investigate
the dynamics of a spinning particle in circular orbit around a Schwarzschild or Kerr black hole in vacuum \cite{Singh2},
or in the presence of radiation, as described by the Vaidya or Kerr-Vaidya metrics \cite{Singh3}.
As mentioned earlier, having an analytic perturbative representation of the MPD equations may become
very useful for identifying the conditions for a transition from stable motion to a chaotic form.
In addition, it may be possible to study the spin-spin interaction between two or more spinning particles,
and determine the possible impact on tidal acceleration effects experienced by a reference particle within this configuration.
There are likely many other applications to follow from future considerations of this formalism.

\section{Conclusion}
\label{sec:5}

This paper has displayed a systematic expression of the Mathisson-Papapetrou-Dixon (MPD) equations in perturbative form that
can be formally extended to infinite order in the perturbation expansion parameter $\varepsilon$, corresponding to the order
of the particle's spin magnitude $s$.
If the spinning extended object in motion is treated as a test particle with a M{\o}ller radius $\rho = s/m \ll r$, where
$r$ is the radius of curvature defined by the gravitational source, then the relevant expressions for the perturbation
will converge rapidly, and only the first- or second-order quantities in $\varepsilon$ are likely required
for most practical calculations.
It may be useful to consider how to extend this formalism for a spinning object that does not satisfy the
``pole-dipole approximation,'' the results of which may perhaps become a suitable barometer for comparing
competing models of extended objects in curved space-time.
This may be a topic worthy of a future investigation.

\begin{acknowledgements}
It is an honour to dedicate this paper to Prof.~Bahram Mashhoon, a much appreciated mentor and friend,
on the occasion of his 60th birthday.
The author is deeply thankful to Profs.~Friedrich Hehl and Claus L\"{a}mmerzahl for inviting him to
contribute to this Festschrift.
He wishes to also thank Prof.~Nader Mobed of the University of Regina for financial and moral support
concerning this project.
The author acknowledges the Prairie Particle Physics Institute at the University of Regina for its hospitality,
where much of this work was completed.
\end{acknowledgements}



\end{document}